\documentstyle[pre,aps,multicol]{revtex} 
\renewcommand{\narrowtext}{\begin{multicols}{2} \global\columnwidth20.5pc}
\renewcommand{\widetext}{\end{multicols} \global\columnwidth42.5pc}

\begin{document}

\draft 

\title{Parametric Ward-Takahashi identity in disordered systems and the
  integral identity associated with the Calogero-Sutherland model}

\author{Nobuhiko Taniguchi}
\address{%
  Department of Physical Electronics, Hiroshima University, Kagamiyama,
  Higashi-Hiroshima 739, Japan}


\maketitle

\begin{abstract}
  By utilizing the symmetric property known as the Ward-Takahashi identity
  in disordered systems, we explore the novel symmetry relations which
  hold in one-dimensional systems with inverse square interaction (the
  Calogero-Sutherland model).  The identities emerge totally from the
  algebraic structure of the model.  They show that the dynamical
  correlators are connected with one another, involving the higher-order
  integrals of motion.  We obtain the result for the coupling strengths
  $\lambda=1/2$, $1$, and $2$, and conjecture that a similar relation may
  hold for arbitrary rational $\lambda$.
\end{abstract}



\narrowtext

\section{Introduction}

The Calogero-Sutherland model (CSM) describes $N$ fermions located on a
ring of the perimeter $L$ with pairwise inverse square
interactions~\cite{Calogero69,Sutherland71}.  The Hamiltonian is given by
\begin{equation}
  H_{\rm CSM} = {1 \over {2m}}\sum_i p_i^2 +{{\hbar ^2\lambda (\lambda
      -1)} \over m}\sum_{i<j} {{\phi ^2} \over {\sin ^2(\phi r_{ij})}},
\label{CSM}
\end{equation}
where $p_i = - i\hbar \partial /\partial r_i$, $r_{ij}=r_i-r_j$, and $\phi
=\pi / L$.  (The usual convention $\hbar=1$ and $m=1/2$ is adopted
hereafter.)
The exact ground state wavefunction of the model is given by a Jastrow
form $\prod_{i<j}\sin^{\lambda}(\phi r_{ij})$, and excited states are
known to be expressed in terms of the symmetric polynomials called the
Jack polynomials~\cite{Forrester94}.
While there is a long history of exactly solvable models in
one-dimensional many-body systems, this model is the only family known so
far where its dynamical correlation functions can be evaluated exactly.
The dynamical density-density correlator $\langle\rho (r,\tau) \rho
(0,0)\rangle$ in the thermodynamic limit was evaluated analytically for
integer and rational values of $\lambda=p/q$~\cite{Lesage95,Ha95}.
A striking simplicity of the result emerges after taking the
thermodynamic limit, which was attained through a lot of mathematical
effort.

The Calogero-Sutherland model is closely connected with the random matrix
theory (RMT), which has successfully been applied to describe the
universal characteristics in quantum chaotic systems, such as compound
nuclei, quantum billiards and quantum dots.  
In the thermodynamic limit, the Jastrow form of the wave function
immediately enables the ground state average to be identified with the
average over Wigner-Dyson ensembles of random matrices for coupling
strengths $\lambda = \beta/2 = 1/2$ (orthogonal), $1$ (unitary), and $2$
(symplectic).
The spectral correlator was generalized to account for spectra that
disperse as a function of some external tunable parameter.  Surprisingly,
it was found that this parametric two-level correlator is identical to the
dynamical density-density correlator of
CSM~\cite{Simons93g,Beenakker93a,Narayan93}.
Though this ``mapping'' between RMT and CSM is available only for the
three special values of the coupling strengths, it can serve as a rich
source of various useful insights in CSM. (See, {\it e.g.},
Ref.~\cite{Taniguchi95b} for recent work in this direction.)

In this paper, we utilize the mapping to examine the symmetric
relation associated with the dynamical correlations in CSM.  To do so, we
extend the Ward-Takahashi identity in disordered systems to incorporate
the parametric correlations with the help of the supermatrix method.
In contrast to the Jack polynomial technique, this method is suitable to
investigate physical quantities in the bulk limit.  It also transparently
provides symmetric properties to dynamical correlation functions, which we
will explore.

\section{Ward-Takahashi identity}

Our starting point is the Ward-Takahashi identity in quantum dots or RMT. 
It asserts that, for the retarded and advanced Green functions
$G^{R,A}_{E}=(E-H \pm i 0)^{-1}$, the identity 
\begin{equation}
{\rm Tr}\left[\:\overline{G^{R}_{E_1} G^{A}_{E_2}}\:\right] = {2\pi 
    i\over \Delta (E_1   - E_2)},
\label{eq:WTI}
\end{equation}
be satisfied in disordered systems, where $\Delta$ is the mean level
spacing and $\overline{\;\cdots\;}$ denotes the averaging over impurity
configurations or random matrices.  Eq.~(\ref{eq:WTI}) results from the
unitarity of the system, so it should be possible to extend this identity
to incorporate the parametric dependence.
Although Eq.~(\ref{eq:WTI}) itself can be proved straightforwardly by
inserting the complete diagonalized basis between $G^{R}$ and $G^{A}$,
such a route of derivation is no longer achieved when they carry different
external parameters, since we can make no common diagonalized basis.
To extend Eq.~(\ref{eq:WTI}) to such situation, we should take account of
the unitarity of the system explicitly.  To do so, we resort to the
supermatrix method~\cite{Efetov83} which translates the unitarity of the
system into the symmetry of the advanced and retarded components.
When we present the results in the context of CSM
(Eqs.~(\ref{eq:I3-identity},\ref{eq:I4-identity}) below), it will be found
that the derived identities take quite simple forms, but still give
nontrivial relations even for the free fermion case ($\lambda=1$).

\section{Summary of the correlators in RMT and CSM}
\subsection{Parametric correlators of RMT}

To make our discussion concrete, take the Hamiltonian
\begin{equation}
H(X) = H_0 + X\Phi,
\end{equation}
where $H_0$ is a random matrix belonging to one of the Dyson ensembles, and
$\Phi$ is a fixed traceless member of the same ensemble.  By use of the
retarded and advanced Green functions
\begin{equation}
G_{E,X}^{R,A}(\bbox{r},\bbox{r}')=\left\langle {\mit{r}}
\right|(E-H(X)\pm i0)^{-1}\left| {\bbox{r}'} \right\rangle,
\end{equation}
we define the following two kinds of universal correlation functions
$k(\omega ,x)$ and $n(\omega ,x)$~\cite{Simons93c}:
\begin{eqnarray}
&&k(\omega ,x)=-{1 \over 2}+{{\Delta ^2} \over {2\pi ^2}}\int\!\!
d\bbox{r}d\bbox{r}'\:\overline{G^R_{1}(\bbox{r},\bbox{r})G^A_{2} 
 (\bbox{r}',\bbox{r}')}\label{eq:def-k}\\%
&&n(\omega ,x)={{s\Delta ^2} \over
 {2\pi^2}}\int\!\! d\bbox{r}d\bbox{r}'\:\overline{
   G^R_{1}(\bbox{r},\bbox{r}')G^A_{2} (\bbox{r}',\bbox{r})}
\label{eq:def-n}
\end{eqnarray}
where the suffix $i=1, 2$ denotes $(E_i,X_i)$ and $s$ is the level
degeneracy which takes account of the Kramers doublets for the
symplectic case.
Rescaled parameters for the energy and the external parameter were
introduced to reveal the universality by
\begin{eqnarray*}
&& \omega \equiv (E_1-E_2)/ \Delta ;\qquad x^2\equiv C(0)(X_1-X_2)^2,\\&&
C(0)=\Delta^{-2}\:\overline{\left( {\partial E_n(X)/ \partial X} \right)^2}.
\end{eqnarray*}

For the orthogonal, unitary, and symplectic ensembles, the analytical
answers for $k(\omega ,x)$ and $n(\omega ,x)$ have been
obtained~\cite{Simons93c,Taniguchi95}.  To present the results
simultaneously for all three ensembles, the integral variables introduced
in Ref.~\cite{Ha95} are convenient.  By assigning $\lambda=p/q=1/2$ for
the orthogonal, $\lambda=p/q=1$ for the unitary, and $\lambda=p/q=2$ for
the symplectic symmetries ($p$ and $q$ are coprimes), they are presented
by
\begin{eqnarray}
  &&k(\omega, x) = {\cal I}\left[ Q^2\, e^{iQ\omega-E x^2/2} \right],
  \\ &&n(\omega, x) = {\cal I}\left[ E\, e^{iQ\omega-E
      x^2/2} \right].
\end{eqnarray}
The integration ${\cal I}[\cdots]$ is defined by
\begin{equation}
  {\cal I}\left[\cdots\right] \equiv C\prod_{i=1}^q 
    {\int_0^\infty\!\!{dx_i}} \prod_{j=1}^p 
    {\int_0^1\!\!{dy_j}}\:F(\lambda |\{x_i,y_j\})\,\left(\cdots\right),
\end{equation}
\begin{mathletters}
\begin{eqnarray}
  &&Q=(2\pi)\left[\sum^{q}_{i=1}x_i + \sum^{p}_{j=1}y_j\right],\\ 
  &&E=(2\pi)^2\left[\sum^{q}_{i=1}\epsilon_{P}(x_i) +
  \sum^{p}_{j=1}\epsilon_{H}(y_j)\right],
\end{eqnarray}
\end{mathletters}%
and $\epsilon_{P}(x)=x(x+\lambda)$ and $\epsilon_{H}(y)=\lambda y
(1-y)$.
The numerical constant $C$ and the form factor $F(\lambda|\{x_i,y_j\})$
were given by
\begin{equation}
C={\lambda ^{2p(q-1)}\;\Gamma ^2(p) \Gamma ^q(\lambda )\Gamma ^p({1\over
    \lambda} ) \over 2\pi^2p!q! \displaystyle{\prod_{i=1}^q \Gamma
    ^2(p-\lambda (i-1))} \displaystyle{\prod_{j=1}^p \Gamma
    ^2(1-\textstyle{j-1\over\lambda})} },
\label{def:C}\\
\end{equation}
\begin{eqnarray}  
F(\lambda |\{x_i,y_j\}) &&= {{\prod_{i<i'} (x_i-x_{i'})^{2\lambda}}
    {\prod_{j<j'} (y_j-y_{j'})^{2/ \lambda }} \over {\prod_{i,j}
      (x_i+\lambda y_j)^2}} \nonumber\\&&\quad\times \prod_{i=1}^q
  \epsilon _P(x_i)^{\lambda-1} \prod_{j=1}^p \epsilon _H(y_j)^{1/
    \lambda-1}.  
\label{def:F}
\end{eqnarray}
In the supermatrix formulation, $F(\lambda|\{x_i,y_j\})$ emerges as a
Jacobian for the integration which is completely determined from the
structure of the underlying graded-symmetric space.

\subsection{Connection with CSM}

The direct connection between the parametric correlations of RMT and
dynamical correlations of CSM is provided when we substitute $\omega \to
r$ and $x^2/2 \to \tau$ ($\tau$ is the Euclidean
time)~\cite{Simons93g,Beenakker93a,Narayan93}.  When we make this
replacement in $k(\omega,x)$, it immediately reproduces the dynamical
density-density correlator $\langle \rho(r,\tau) \rho(0,0)\rangle$ for
$\lambda = 1/2$, $1$, and $2$, i.e.,
\begin{equation}
  \left\langle \rho (r,\tau)\rho (0,0) \right\rangle
    ={\cal I}\left[Q^2 \cos (Qr)\, e^{-E\tau}\right].
\end{equation}
The other function $n(\omega, x)$ is found to be related to the dynamical
current-current correlator of CSM~\cite{Taniguchi95b},
\begin{equation}
  \left\langle j(r,\tau)j(0,0) \right\rangle
    ={\cal I}\left[E^2 \cos (Qr)\, e^{-E\tau}\right].
\end{equation}
Since the Ward-Takahashi identity Eq.~(\ref{eq:WTI}) states ${\cal I}[E\,
e^{iQ\omega}] = - 1/(i\pi\omega)$, it characterizes the current-current
correlator of CSM rather than the density-density correlator.

\section{Derivations and Results}

Now we present how we can extend and derive the Ward-Takahashi identity to
the case for finite $x$, or dynamical correlations.  To avoid the
notational confusion, we use $(\omega,x)$ of RMT, instead $(r,\tau)$ of
CSM, but by substituting $\omega \to r$ and $x^2/2 \to \tau$, we can
obtain the corresponding expressions for CSM on each step.
We follow Refs.~\cite{Verbaarschot85b,Zuk94} to derive the
Ward-Takahashi identity within the framework of the supermatrix method.
The basic underlying idea is to translate the unitarity of the system into
the hyperbolic symmetry between the advanced and retarded components.
(See Eq. (\ref{eq:deltaT}) below.)

Consider the generating function for $k(\omega, x)$ and $n(\omega,
x)$ in the supermatrix nonlinear-$\sigma$ model formulation~\cite{Efetov83},
\begin{equation}
  Z_J=\left\langle {\exp \left[ {{\rm STr}\left( {{\cal Q}J} \right)} \right]}
\right\rangle _{\cal Q},
\end{equation}
where ${\cal Q}$ is the $8\times 8$ supermatrix satisfying ${\cal Q}^2=1$
and its explicit structure of ${\cal Q}$ can be found in Ref.~\cite{Efetov83}.
The supertrace ${\rm STr}$ is defined by ${\rm STr}(\cdots) = {\rm
  Tr}[(k_F-k_B)(\cdots)]$ where $k_{\alpha}$ is a projector either onto
the Bose space ($\alpha=B$) or onto the Fermi space ($\alpha=F$).  The
source matrix $J$ is chosen as ($a$ and $b$ are $c$-numbers)
\begin{equation}
  J= (a + b\Sigma_1) \Lambda k_{\alpha}. 
\end{equation}
Note that we are allowed to use either one to generate $k(\omega,x)$
and $n(\omega,x)$.  $\Lambda$ and $\Sigma_1$ are $8\times 8$ matrices
defined by
\begin{equation}
  \Lambda = \left(
    \begin{array}{cc} 1_4 & 0 \\ 0 & -1_4
    \end{array}\right);\qquad  
  \Sigma_1 = \left(
    \begin{array}{cc} 0 & 1_4 \\ 1_4 & 0 \end{array}\right) 
\end{equation}
The average $\langle \cdots \rangle_{\cal Q}$ denotes the integral $\int\!
D{\cal Q} \left( \cdots \right) e^{-F[{\cal Q}]}$.  Corresponding to
$\lambda=p/q=1/2$ (orthogonal), $1$ (unitary), and $2$ (symplectic),
$F[{\cal Q}]$ is equal to
\begin{equation}
  F[{\cal Q}]=p\;\left\{ {i\pi \omega\over 4} {\rm STr}({\cal Q}\Lambda
    )-{\lambda \pi^2 x^2\over 16} {\rm STr}({\cal Q}\Lambda)^2\right\}. 
\end{equation}
When we make the infinitesimal rotation on the saddle-point manifold as
${\cal Q}\to {\cal Q}'=(1-\delta T){\cal Q}(1+\delta T)$, $Z_J$ remains
invariant because of the integration over the manifold $Q$.  Hence
\begin{eqnarray}
  &&\delta Z_J=\left\langle e^{{\rm STr}{\cal Q}J}\:
{\rm STr}\left[ \left(
[J,{\cal Q}] - {p i\pi \omega\over 4} [\Lambda,{\cal Q}]\right.\right.\right.
\nonumber \\ 
&&\qquad \left.\left.\left. + {p^2\pi^2 x^2\over 8q}\left\{ 
(\Lambda {\cal Q})^2 - ({\cal Q}\Lambda)^2\right\}
\right)\delta T\right]  \right\rangle_{\cal Q} =0.
\end{eqnarray}
Although this identity holds for arbitrary infinitesimal rotations $\delta
T$, we particularly choose ($\alpha=B$ or $F$)
\begin{equation}
  \delta T \propto k_{\alpha} \Sigma_1.
\label{eq:deltaT}
\end{equation}
The choice reflects $U(1,1)$ symmetry of the advanced and retarded
components within the bosonic or fermionic sector.  This hyperbolic
symmetry is responsible for producing the Ward-Takahashi identity in the
supermatrix method.
After substituting Eq.~(\ref{eq:deltaT}) for $\delta T$, we have
\begin{equation}
  \left\langle \left(b q_1+{4a - p i\pi \omega  \over 4}
     q_2 + {p^2\pi ^2x^2 \over 8q} q_3\right) e^{aq_1+bq_2} \right\rangle_{Q}
 = 0,
\label{eq:gf-of-ab}
\end{equation}
where we define
\begin{mathletters}
\begin{eqnarray}
  &&q_1 = {\rm STr}\left[ k_{\alpha}\Lambda {\cal Q} \right],\\
  &&q_2 = {\rm STr}\left[ k_{\alpha}\Sigma_1\Lambda {\cal Q}\right],\\
  &&q_3 = {\rm STr}\left[ k_{\alpha}\Sigma_1 (\Lambda {\cal Q})^2 \right].
\end{eqnarray}
\end{mathletters}
Note that the correlator $k(\omega,x)$ and $n(\omega,x)$ are related by
\begin{mathletters}
\begin{eqnarray}
  &&{1 \over 16 } \langle (q_{1})^2 \rangle_{\cal Q} = 1 + k(\omega,x)
  \label{eq:q1q1} \\
  &&{1 \over 16} \langle (q_{2})^2 \rangle_{\cal Q} = \pm n(\omega,x)
\label{eq:q2q2}
\end{eqnarray}
\end{mathletters}
Depending on the choice of $k_B$ or $k_F$, we have positive or
negative sign in front of $n(\omega,x)$.  

From Eq.~(\ref{eq:gf-of-ab}), we can readily derive a sequence of integral
identities by comparing each coefficient of polynomials of $a$ and $b$.
Not all of them, however, produce nontrivial integral identities.
We can show that the coefficients of $a^0 b^0$ and $a^1 b^0$ vanish
trivially.  The first nontrivial identity comes from the coefficient of
$a^0 b^1$, i.e.,
\begin{equation}
\frac{1}{p} \left\langle {q_1} \right\rangle _{\cal Q}=\left\langle {{i\pi
      \omega} \over 4}{(q_2)^2} -{{\lambda^2\pi ^2x^2} \over
    {8}}{q_2q_3}\right\rangle_{\cal Q}.    
\end{equation}
After some straightforward but rather lengthy evaluation of the
supermatrix integration for all three values of $\lambda$, we obtain the
result which can be summarized as follows (restoring $\omega\to r$ and
$x^2/2\to \tau$):
\begin{equation}
  \frac{1}{\pi} = {\cal I}\left[\left({-i r E+\tau I_3}\right)
  e^{iQr-E\tau}\right],
\label{eq:I3-identity}
\end{equation}
where we define 
\begin{eqnarray}
  I_n &&\equiv (2\pi)^n\left[ \sum_{i=1}^{q} {x_i(x_i+\lambda)
      (2x_i+\lambda)^{n-2}} %
\right. \nonumber \\ &&\qquad \qquad \left. %
+\lambda ^{n-1}\sum_{j=1}^{p}
  {y_j(1-y_j)(1-2y_j)^{n-2}}\right].
\label{eq:def-In}
\end{eqnarray}
Eq.~(\ref{eq:I3-identity}) serves as the extension of the Ward-Takahashi
identity Eq.~(\ref{eq:WTI}), and consists of the main result of the
paper.  


We can go on to the higher-order identity from Eq.~(\ref{eq:gf-of-ab}), on
principle, but the evaluation of the integration becomes harder and harder
to complete.  Among the second-order polynomials of $a$ and $b$, we can
confirm that only the coefficient of $a^{1}b^{1}$ gives the nontrivial
relation:
\begin{equation}
\frac{1}{p} \left\langle {(q_1)^2+(q_2)^2} \right\rangle _{\cal
  Q}=\left\langle {{i\pi\omega } \over 4} {q_1(q_2)^2} - {{\lambda^2\pi
      ^2x^2} \over {8}}{q_1q_2q_3} \right\rangle _{\cal Q}. 
\end{equation}
However, as we see from Eq.~(\ref{eq:q2q2}), this will depend on the
values of $\lambda$ ($p$ and $q$) as well as the choice of $\alpha=B$ or
$F$.  Hence for each value of $\lambda$, we have two integral identities.
From these, we can seek an interesting form of the identity which seems
the direct extension of Eq.~(\ref{eq:I3-identity}), which can be presented by
\begin{eqnarray}
  &&1 + {\cal I}\left[ (Q^2 +\frac{q-p}{p} E)\; e^{iQr-E\tau}\right]
  \nonumber \\ &&\qquad = \frac{1}{p^2}\, {\cal I}\left[ \left(-i r I_3 +
  \tau I_4 \right) e^{iQr - E\tau}\right].
\label{eq:I4-identity}
\end{eqnarray}
Note that the coefficients of Eq.~(\ref{eq:I4-identity}) are deduced to
reproduce the actual results of $\lambda=1/2$, $1$, and $2$.

\section{Discussion}

There are known multiple integral identities which are associated with
CSM.  They are called the Selberg integrals~\cite{Mehta}, and their
generalization by Dotsenko and Fateev~\cite{Dotsenko85} are particularly
useful.  For instance, they were used to determine the correct
normalization factor of the correlation functions~\cite{Forrester95}.  We
also mention that the Dotsenko-Fateev integral can provide a systematic
means to evaluate a certain correlation function which showed up in
disordered systems~\cite{Forrester96}.
However, we emphasize that those integral formulae are not powerful enough
to explain Eq.~(\ref{eq:I3-identity}), because they can be applied only
when the integrands are polynomials.  The simple form of the identity
Eq.~(\ref{eq:I3-identity}) may suggest that these known multiple integral
formulae be extended somehow for the case involving an exponential factor
such as $e^{iQr-E\tau}$.
We remark that the derived integral identity Eq.~(\ref{eq:I3-identity}) is
not trivial at all from the mathematical point of view, even for the
simplest case of the free fermion ($\lambda=1$), though we can convince
ourselves of its correctness, {\it e.g.}, by checking the asymptotics, or
evaluating for small $\tau$ expansion.

The quantities $I_n$ (for $n\ge 3$) correspond to the higher-order
integrals of motion of CSM, as well as $Q$ and $I_2=E$.  To see this
transparently, identify the velocities $v_i$ ($\bar v_j$) for particles
(holes) by
\begin{mathletters}
\begin{eqnarray}
  && v_i = v_s (1+2x_i/\lambda),\\
  && \bar v_j = v_s (1-2y_j),
\end{eqnarray}
\end{mathletters}%
where $v_s = \pi\lambda\hbar\rho_0/m = 2\pi\lambda$ is the sound
velocity~\cite{Zirnbauer95}. 
Since the velocity (rapidity) is the conserved quantity of CSM, 
\begin{equation}
  J_n = m\sum_{i=1}^{q} v_i^n + m_h \sum_{j=1}^{p} \bar v_j^n,
\end{equation}
should act as the integrals of motion, so does $I_n = J_n/2 - 2\pi^2
\lambda^2 J_{n-2}$.  
In Ref.~\cite{Romer96}, a few kinds of the higher-order integrals of
motion were investigated.  Although the similarity of their look, the
direct connection with $I_n$ in Eq.~(\ref{eq:def-In}) is missing at
present.

\section{Conclusion}

In conclusion, we have derived the Ward-Takahashi identity for the
parametric correlations of RMT.  By doing so, it was shown that there
exist novel integral identities which are associated with the
dynamical correlations of CSM. It is remarked that they amount to a
new generalization of the Selberg integration. 
As was seen from the derivation in the context of RMT, this is the
manifestation of the unitarity of the system, i.e., the hyperbolic
symmetry of the advanced and retarded components.  However, its nature and
implication in CSM is not so clear at present.
Since our arguments rely heavily on the mapping between RMT and CSM, we
can make no decisive statement on the validity of the derived integral
identities for {\em arbitrary rational}\/ values of $\lambda$.  We can,
however, suggest two possible scenarios:
Eqs.~(\ref{eq:I3-identity},\ref{eq:I4-identity}) are (1) true only for
$\lambda=1/2, 1, 2$, or (2) true for all rational values of $\lambda$.  If
the latter were true, it would remain as a future challenge how the
integral identities Eqs.~(\ref{eq:I3-identity},\ref{eq:I4-identity}) can
be deduced from the Jack polynomials, or the $W$-algebra which is known as
the symmetry of CSM~\cite{Hikami93}.

\section*{Acknowledgments}
This work was initiated through the discussion with B. D. Simons.  The
author is also grateful to B. L. Altshuler, P. J. Forrester, J. Kaneko and
B. S.  Shastry for their interest and useful discussions.  This work was
supported in part by Grant-in-Aid for Scientific Research No.~08740247
from the Ministry of Education, Science, Sports and Culture of Japan.

%

\widetext

\end{document}